\documentclass[aip,reprint,onecolumn,floatfix]{revtex4-2}
\usepackage{graphicx}
\usepackage{amsfonts}
\usepackage{amsmath}
\usepackage{rotating}
\usepackage{amssymb,amsthm}
\usepackage{xcolor}
\usepackage[normalem]{ulem}
\usepackage{nicefrac}

\theoremstyle{definition}

\theoremstyle{remark}

\newcommand{\mbold}[1]{\mbox{\boldmath{\ensuremath{#1}}}}

\def\be{\begin{equation}}
\def\ee{\end{equation}}
\def\beq{\begin{eqnarray}}
\def\eeq{\end{eqnarray}}
\def \Re {\mbox{Re}}
\def \Im {\mbox{Im}}

\def \bell {\mbox{{\mbold\ell}}}
\def \bk {\mbox{{\bf k}}}

\def \bm {\mbox{{\bf m}}}
\def \bh {\mbox{{\bf h}}}
\def \bomega {\mbox{{\mbold \omega}}}

\begin{document}

\title{Spatially Homogeneous Teleparallel Gravity: Bianchi I}

\author {A. A. Coley}
\email{aac@mathstat.dal.ca}
\affiliation{Department of Mathematics and Statistics, Dalhousie University, Halifax, Nova Scotia, Canada, B3H 3J5}

\author{R. J. {van den Hoogen}}
\email{rvandenh@stfx.ca}
\affiliation{Department of Mathematics and Statistics, St. Francis Xavier University, Antigonish, Nova Scotia, Canada, B2G 2W5}

\date{\today}

\begin{abstract}

Using a recently developed algorithm that chooses preferred coordinates and a preferred co-frame, we will determine the completely general Bianchi type I teleparallel geometry.  In using  this algorithm, any remaining gauge freedom is allocated to the choice of spin connection.  We then solve the symmetry constraints placed on the spin connection to derive a general class of Bianchi type I teleparallel geometries.  We find that this general class of Bianchi type I teleparallel geometries breaks naturally into two sub-classes.  We then illustrate some simple power-law solutions in $F(T)$ teleparallel gravity for each case to illustrate the differences.

\end{abstract}

\maketitle


\section{Introduction}

The six-parameter $\Lambda$CDM model within the theory of General Relativity (GR) is consistent, up to some tensions, with current cosmological and astrophysical observations \cite{Planck:2018vyg,WMAP:2012nax}. The tensions include the inexplicable difference in the measured values of the current Hubble parameter using two independent techniques (see \cite{Riess:2016jrr,Ananthaswamy:2018ohs,DiValentino:2022fjm,Hu:2023jqc} for details).  The $\Lambda$CDM model assumes that the geometry of the observable universe is flat and the dark energy (the source of our currently observed acceleration of the universe) is consistent with a cosmological constant. However, the nature of the dark energy continues to elude us.  Given that it accounts for nearly 70\% of the total matter-energy budget of the universe, one might think we could develop a better way to explain this current acceleration. Modified theories of gravity \cite{Clifton:2011jh,Nojiri_Odintsov2006,Capozziello_DeLaurentis_2011} could provide either partially or wholly an alternative explanation for the currently observed acceleration of the universe.

There is continued interest in a class of modified gravitational theories in which the dynamics of the gravitational field are encoded in the torsion instead of the curvature \cite{Obukhov_Pereira2003,Aldrovandi_Pereira2013,Ferraro:2006jd,Ferraro:2011us,Cai_2015,Krssak2015,Bahamonde_Boehmer_Wright2015,Krssak:2018ywd,Bahamonde:2021gfp}. This includes teleparallel gravity in which the curvature and nonmetricity are identically zero.  Interestingly, there is a subclass of teleparallel gravity that is dynamically equivalent to GR \cite{Obukhov_Pereira2003,Aldrovandi_Pereira2013}. The Lagrangian for this teleparallel equivalent to General Relativity (TEGR) theory is a particular scalar quantity,
$T$, constructed from the torsion which differs from the Lagrangian of GR by a total derivative. Therefore the field equations for TEGR are formally equivalent to GR locally. Given that GR passes all current experimental solar system tests, one might expect that any divergence from GR or TEGR would be small at the present time.

One popular generalization of TEGR is $F(T)$ gravity \cite{Ferraro:2006jd,Ferraro:2011us,Cai_2015,Krssak2015,Bahamonde_Boehmer_Wright2015,Krssak:2018ywd,Bahamonde:2021gfp}. If the teleparallel geometry is defined in a gauge invariant manner as a geometry with zero curvature, then the most general spin-connection that satisfies this requirement is the purely inertial spin-connection, which vanishes in a very special class of frames (“proper frames”) where
all inertial effects are absent, and non-zero in all other frames  \cite{Obukhov_rubilar2006,Lucas_Obukhov_Pereira2009,Aldrovandi_Pereira2013,Krssak:2018ywd}. The primary
benefit of this covariant approach is that, by using the purely inertial connection, the resulting teleparallel gravity theory is locally Lorentz invariant \cite{Krssak_Saridakis2015,Krssak:2018ywd}.  We adopt the covariant approach here.
We note, it is then possible to use this arbitrary co-frame in an arbitrary coordinate system with the corresponding spin-connection to produce equivalent field equations to those in the
proper frame \cite{Krssak_Saridakis2015,Krssak_Pereira2015}; that is, we can always transform the analysis to the ``Pure Tetrad'' approach to teleparallel gravity.

In teleparallel gravity, the frame (or co-frame) basis  together with a spin connection replaces the metric as the primary objects of study. Therefore the situation is complicated in that in addition to an initial ansatz for the co-frame we also require an initial ansatz for the spin connection.  In some situations, one commonly chooses a diagonal co-frame and a trivial spin connection.  This choice does indeed work in some instances; for example, the zero curvature Robertson-Walker or Bianchi I in Cartesian coordinates can be analyzed this way without loss of generality.  However, in general a diagonal co-frame ansatz will require a non-trivial spin-connection or alternatively, a non-diagonal co-frame can be employed with a trivial spin connection. See \cite{Coley:2022qug} for examples in spatially curved teleparallel Robertson-Walker (TRW) geometries. For a particular teleparallel theory of gravity, to determine a solution of the field equations one must first choose a coordinate system, $x^\mu$, a co-frame basis, $\bh^a$, and a spin-connection, $\bomega^a_{~b}$  (or, alternatively, a coordinate system and a proper frame basis in which the spin-connection is trivial). We will call the solutions of the field equations to teleparallel gravity theories, {\em{teleparallel geometries}}.

In GR, when looking for solutions to the field equations, one typically assumes some symmetry that reduces the complexity of the resulting equations.  Using these symmetries in GR is a common practice and well-established.  However, in teleparallel gravity, since we require ansatzes for two different quantities, the co-frame and the spin-connection, our ansatzes, in the presence of any symmetry must be consistent \cite{Bahamonde:2021gfp}. Care must be taken when selecting ansatzes for the co-frame spin-connection pair, else the situation immediately reduces to TEGR (or GR) with a cosmological constant and a re-scaled coupling constant \cite{Krssak:2018ywd,Bahamonde:2021gfp}.

Similar to the challenges in GR, it is possible that two seemingly distinct choices of the coordinates, co-frame basis and spin-connection which satisfy the teleparallel field equations
are, in fact, the same solution.  However this equivalence may be hidden by our choices. To determine the equivalence of two teleparallel geometries in an invariant manner, a
modification of the Cartan-Karlhede algorithm adapted to Riemann-Cartan geometries has been introduced \cite{Coley:2019zld}. For example, it has been shown that inequivalent solutions to the teleparallel field equations can give
equivalent metrics \cite{Coley:2019zld}. This suggests that for a given co-frame/spin-connection anzatz we can produce distinct teleparallel geometries by choosing different spin-connections which cannot be related to
each other using local Lorentz frame transformations. This observation motivates the search for other spin-connections which generate distinct torsion geometries with the same metric geometry. Here, in the current work, our target metric geometry will be Bianchi I, however we shall assume a completely general spin-connection that is consistent with the space-time symmetries.

It is well-known that Bianchi I geometries provide the simplest anisotropic solutions in GR.  Further, the vacuum Bianchi I solutions in GR (i.e., Kasner solutions) play an extremely important role in the past evolution of cosmological models in GR.  It is strongly believed (the BKL conjecture) that typical past behavior of anisotropic cosmological models in GR approaches that of an infinite sequence of different vacuum Kasner solutions \cite{Hobill:1993osv,ellis_uggla_wainwright_1997,hobill_1997,Coley:2003mj}. Indeed, there is growing evidence that this infinite sequence of Kasner solutions is also the typical past attractor for large classes of inhomogeneous cosmological models \cite{Berger:2002st,Uggla:2003fp}. Due to the difficulty in constructing consistent anisotropic cosmological models in teleparallel theories of gravity, it is unknown whether there is similar behaviour present.  One of the purposes of this work is to construct the most general Bianchi I geometry consistent with $F(T)$ teleparallel gravity.

\subsection{Review of other work}

There has been some previous analysis of spatially homogeneous but anisotropic cosmological models in $F(T)$ teleparallel gravity in the literature
\cite{Rodrigues:2012qua,Rodrigues:2014xam,Cai_2015,Bahamonde:2021gfp,Paliathanasis:2022vux,
      Bahamonde:2021gfp,Amir:2015wja,Paliathanasis:2022vux,Sharif:2011bi,
      Fayaz:2014swa,Fayaz:2015yka,
      Skugoreva:2017vde,Skugoreva:2019bwt,Tretyakov:2021cgb,Rodrigues:2013iua,Aslam:2013coa,Paliathanasis:2016vsw,Paliathanasis:2017htk}. The most common spatially homogeneous but anisotropic models studied in the literature have been Bianchi types I, III and Kantowski Sachs (KS).

Bianchi type III anisotropic cosmological models in $F(T)$ teleparallel gravity have been studied in \cite{Rodrigues:2012qua,Rodrigues:2014xam,Cai_2015,Bahamonde:2021gfp,Paliathanasis:2022vux}.  In \cite{Rodrigues:2012qua,Rodrigues:2014xam,Cai_2015} the authors assumed a diagonal co-frame and trivial spin connection.  With these restrictive assumptions, one of the field equations immediately reduced to the equivalent of $\partial_t(F'(T))=0$.  This implies that the $F(T)$ theory is equivalent to TEGR (or GR) with a cosmological constant and a re-scaled matter coupling constant.  In essence, the assumption of a diagonal co-frame with a trivial spin connection is inconsistent with the $F(T)$ field equations.  Either the spin connection should be non-trivial, or the co-frame be non-diagonal.  Fortunately, consistent ansatzes (ansatzes which do not immediately lead to TEGR) for Bianchi III geometries have been put forward in \cite{Bahamonde:2021gfp,Paliathanasis:2022vux}, where it is shown that a proper co-frame necessarily requires a that the co-frame ansatz be non-diagonal.  Indeed, proper co-frame anstazes for all Bianchi types are provided in \cite{Bahamonde:2021gfp} in which the spin connection is trivial. However, it is not clear whether these proper co-frame ansatzes in \cite{Bahamonde:2021gfp} represent special cases, or can be used to represent the most general Bianchi type teleparallel geometries.

Kantowski-Sachs anisotropic models also have been developed in \cite{Cai_2015,Rodrigues:2012qua,Rodrigues:2014xam,Amir:2015wja,Bahamonde:2021gfp,Paliathanasis:2022vux}. Unfortunately the metric ansatz describing the KS geometries in \cite{Rodrigues:2012qua,Cai_2015,Bahamonde:2021gfp} was incorrect. In an updated paper the authors of \cite{Rodrigues:2012qua}  have seemingly corrected the situation \cite{Rodrigues:2014xam}. However, in both \cite{Rodrigues:2014xam,Amir:2015wja} which have correct ansatzes for the co-frame, the authors assumed a diagonal co-frame with a trivial spin-connection. These assumption once put into the $F(T)$ teleparallel field equations results in TEGR (or GR) with a cosmological constant and a re-scaled matter coupling constant. The correct co-frame spin-connection pair for KS geometries does appear in \cite{Paliathanasis:2022vux}.  In \cite{Paliathanasis:2022vux}, the author coupled TEGR to a scalar field in an effort to find some exact solutions.

Fortunately, for Bianchi type I geometries, a diagonal co-frame with a trivial spin connection is always consistent with the $F(T)$ teleparallel field equations. In many studies one tries to reconstruct $F(T)$ based on some external assumptions on the matter source, for example, dust or radiation, \cite{Rodrigues:2012qua,Rodrigues:2014xam,Sharif:2011bi}.  In \cite{Rodrigues:2012qua} the authors looked for and found de Sitter and power-law type solutions to the field equations. Building on the analysis in \cite{Rodrigues:2012qua}, it was shown that anisotropic Bianchi type I geometries isotropize to the future \cite{Rodrigues:2014xam}, but it was assumed that $F(T)=T-2\Lambda$, which is just TEGR with a cosmological constant.  In \cite{Sharif:2011bi}, in addition to reconstruction methods for $F(T)$  based on assumed equations of state for the matter, the authors undertook analysis to determine the equation of state where $F(T)$ is assumed apriori. In a pair of papers \cite{Fayaz:2014swa,Fayaz:2015yka} the authors determined de Sitter and power-law solutions.  They also investigated the possibility of constructing $F(T)$ from both holographic dark energy (HDE) and Ricci dark energy (RDE) arguments.

In \cite{Skugoreva:2017vde}, the authors built Kasner Bianchi type I solutions in $F(T)$ teleparallel gravity.  They assumed $F(T)=T+\alpha T^n$ with no matter source, in which case the field equations immediately impose the constraint that $T$ is a constant.  The standard GR Kasner solution and de Sitter solution are shown to exist with the assumed form for $F(T)$.  Further, in \cite{Skugoreva:2017vde} they showed through numerical computations that typical future behaviour in Bianchi type I models is the isotropic de Sitter solution while the past behaviour is a Kasner power-law solution. The authors of \cite{Skugoreva:2017vde} expanded upon their analysis to include a perfect fluid source \cite{Skugoreva:2019bwt}. With the addition of matter, using the same $F(T)$, the authors of \cite{Skugoreva:2019bwt} concluded that the matter content of the universe plays an important and non-trivial role in the past behaviour near the cosmological singularity, contrary to the BKL conjecture of GR.  A slightly different approach to determining the behaviours in Bianchi I geometries containing a perfect fluid  was undertaken in \cite{Tretyakov:2021cgb},  in which the solution for the torsion scalar, Hubble scalar, and the energy density is assumed to all have power-law forms.  In the case $F(T)=T+\alpha T^2$, it was shown that isotropic de Sitter solution is stable to the future in an expanding universe if $\alpha >0$; that is, the anisotropic model isotropizes.

In \cite{Paliathanasis:2016vsw}, the authors investigated Bianchi type I models where it was assumed that $F(T)=T^n$. The the co-frame was assumed to be diagonal and the three co-frame functions were power-law in nature $a_i(t)=a_{i0}t^{p_i}$. They found that two Kasner-like possibilities could occur:
\begin{itemize}
\item $p_1=p_2=p_3$ and $n = 1/2$: Here $F(T)=\sqrt{T}$. The solution represents an isotropic $k=0$ TRW geometry.
\item $p_1+p_2+p_3 = 2n-1,\ p_1^2+p_2^2+p_3^2=(2n-1)^2$: In this case it was possible to re-scale to $\bar{p}_i=p_i/(2n-1)$ so that one obtains the usual Kasner conditions. The Kasner solution only occurs when $n=1$.  Also, if $n>0$ the chaotic dynamical behaviour on approach to the cosmological singularity in the past, via an infinite sequence of Kasner eras, can occur as in GR.
\end{itemize}
Further, the authors in \cite{Paliathanasis:2016vsw} carefully considered the form and nature of the ``$00$'' vacuum field equation. They argued that one possibility to satisfy this particular field equation is to have $T=0$ and require $\lim_{T\to 0}F(T)=0$. Otherwise, one must require $F(T)=\sqrt{T}$, which does not have GR in the appropriate limit.
Building upon the results in \cite{Paliathanasis:2016vsw}, the stability of the Kasner solutions were investigated in \cite{Paliathanasis:2017htk}. In \cite{Paliathanasis:2017htk} the authors assumed that $F(T)=T+\alpha T^n$ and $F(T)=T+\alpha(1-e^{-pT})$.  In both cases the authors concluded that the Kasner geometry is a saddle point, which means it is unstable to the past in $F(T)$ teleparallel gravity for the assumed forms of $F(T)$ employed.  The authors also demonstrated that Bianchi I models isotropized to the de Sitter solution to the future \cite{Paliathanasis:2017htk}.

Clearly, more work needs to be done in understanding the nature of Bianchi I geometries in $F(T)$ teleparallel gravity. Without this solid base, it becomes very difficult to move onto more general classes of spatially homogeneous and anisotropic models.  One of our goals here is to determine the most general Bianchi I geometry in $F(T)$ teleparallel gravity, and to construct where possible reasonable solutions, and to illustrate that different solutions for the co-frame will yield the same metric.

\subsection{Notation}

We assume a differentiable manifold $M$ with coordinates $x^\mu$.  We define $\bh^a = h^a_{~\mu}\,dx^\mu$ to be the basis one forms with $\bh_a = h_a^{~\mu}\partial_\mu$ representing the dual vector basis.  The spin connection one-form is expressed as $\bomega^a_{~b}=\omega^a_{~bc}\,\bh^c$ which is assumed to be metric compatible and have zero curvature. We denote local Lorentz transformations of the co-frame basis as $\bh'^a=\Lambda^a_{\phantom{a}b}\bh^b$, and define $\Lambda_b^{\phantom{a}a} \equiv (\Lambda^{-1})^a_{\phantom{a}b}$. The first and second derivatives of the function $F(T)$ are denoted as $F_T(T)$ or $F'(T)$ and $F_{TT}(T)$ or $F''(T)$ respectively.


\section{Review and Setup}

\subsection{Symmetry}

An isometry, which is a diffeomorphism from a geometrical space into itself which preserves the metric, is well understood and extensively utilized in metric based theories of gravity such as GR.  In theories of gravity in which the metric is not the primary object of interest, for example in Teleparallel theories of gravity, the role of symmetries is not as clear.

Assuming that the group of symmetries has a trivial isotropy subgroup, we can define an affine frame symmetry as a diffeomorphism  on $\mathcal{F}(M)$, the frame bundle of the manifold $M$ \cite{olver1995}, such that there exists a frame in which the
vector field, ${\bf X}$, satisfies \cite{Fonseca-Neto:1992xln}:
\beq
\mathcal{L}_{{\bf X}} \bh_a = 0 \text{ and } \mathcal{L}_{{\bf X}} \bomega^a_{~bc} = 0 \label{Intro:FS2}
\eeq
where $\bf X$ is the vector field generated by the affine symmetry.  This definition is {\it a frame-dependent} analogue of the definition of a symmetry introduced by \cite{HJKP2018,Pfeifer:2022txm,Hohmann:2015pva}.

The affine symmetries of a particular teleparallel geometry may not coincide with the set of
isometries for the metric. An affine frame symmetry is an isometry (i.e., ${\bf X}$ is a Killing vector field --
$\mathcal{L}_{{\bf X}} g_{\mu\nu} = 0$), but not all isometries are affine frame symmetries  \cite{Coley:2019zld}.
Indeed, if $\bf X$ is an isometry then $\mathcal{L}_{{\bf X}} \bh^a = \lambda^a_{~b}\bh^b$ for some $\lambda^a_{~b}\in \mathfrak{so}(1,3)$ which certainly could be the zero matrix, but need not be. If the isometry group of the metric has a trivial isotropy subgroup, then we can always find an invariant frame $\bh^a$ such that  $\mathcal{L}_{{\bf X}} \bh^a =0$ satisfying our definition above.  However, when the symmetry group of the metric has a non-trivial isotropy group, then another procedure is required to produce a invariant frame \cite{McNutt_Coley_vdH2023} that can be employed in such situations.

\subsection{Lorentz Transformations}

We shall adopt the {\bf Complex Null Gauge} with complex null frame $ \{ \bh {^a} \} = \{\bk, \bell, \bm, \bar{\bm} \}$ \cite{classc}, in which
\beq
g_{ab} = \left[ \begin{array}{cccc} 0 & -1 & 0 & 0 \\ -1 & 0 & 0 & 0 \\ 0 & 0 & 0 & 1 \\ 0 & 0 & 1 & 0 \end{array}\right].
\label{gauge}
\eeq
We note that $g_{ab}$ is invariant under the proper ortho-chronous Lorentz group, $SO^+(1,3)$. These Lorentz transformations $\Lambda^a_{~b}$ are local functions of the coordinates and transform the metric, coframe and spin connection as
\begin{eqnarray}
g_{ab}         &\to& {g'}_{ab}        =(\Lambda^{-1})^c_{~a}(\Lambda^{-1})^d_{~b}g_{cd} = g_{ab},\\
\bh^a          &\to& {\bh'}^a         =\Lambda^a_{~b} {\bh}^b,\\
\bomega^a_{~b} &\to& {\bomega'}^a_{~b}=\Lambda^a_{~c} \bomega^c_{~d}(\Lambda^{-1})^c_{~b} + \Lambda^a_{~c} \,d(\Lambda^{-1})^c_{~b}.
\end{eqnarray}

Most physical theories of gravity are assumed to be invariant under general linear transformations (the {\emph{Principle of Relativity}}).  Therefore, the corresponding field equations must transform homogeneously.  Since we have made a gauge choice, the field equations must be invariant under the remaining gauge freedom, in this case $SO^+(1,3)$ Lorentz transformations. In our chosen null gauge, the Lorentz frame transformations can be elegantly represented \cite{Stephani} using two complex valued functions $B(x^\mu)$ and $E(x^\mu)$, parameterizing the null rotations and a pair of real valued functions $A(x^\mu)$ and $\theta(x^\mu)$ parameterizing the boosts and spins (spatial rotations), respectively. The Lorentz transformations are explicitly given in \cite{Stephani,Coley:2022aty}.

A composition of the four different possible transformations yields
\beq
\Lambda^{a}_{~b}= \begin{bmatrix}
A & A^{-1}E\bar{E} & \bar{E}e^{i\theta} & Ee^{-i\theta}  \\
AB\bar{B}\  & A^{-1}(1+B\bar{B}E\bar{E}+\bar{B}E+B\bar{E})\  & (B\bar{B}\bar{E}+\bar{B})e^{i\theta}\  & (B\bar{B}E+B)e^{-i\theta}   \\
AB & A^{-1}(BE\bar{E}+E) & (1+B\bar{E})e^{i\theta} & BEe^{-i\theta}  \\
A\bar{B} & A^{-1}(\bar{B}E\bar{E}+\bar{E}) & \bar{B}\bar{E}e^{i\theta} & (1+\bar{B}E)e^{-i\theta} \\
\end{bmatrix}
\label{general_lorentz}
\eeq
where the inverse is also displayed in \cite{Coley:2022aty}. This general Lorentz transformations has six degrees of freedom, and a repeated Lorentz transformation only simply serves to redefine
the Lorentz parameter functions $\{A, \theta, B, E\}$.

\subsection{Spin connection}

With the assumption of a {\bf Complex Null Gauge}, then the most general Lorentz transformation can be expressed as given in equation \eqref{general_lorentz}. Therefore, the coordinates of the most general spin connection one-form $\omega^a_{~b\mu} \equiv (\Lambda^{-1})^{a}_{~b}\partial_\mu(\Lambda^b_{~c})$ computed from this Lorentz transformation can be expressed elegantly as \cite{Coley:2022aty}:
\beq
\omega^a_{~b\mu}
=\begin{bmatrix}
Re(\Theta_\mu)& 0 & \Psi^{I}_{~\mu} & \bar{\Psi}^{I}_{~\mu}  \\
0 & -Re(\Theta_\mu) & \bar{\Psi}^{II}_{~~\mu}  & \Psi^{II}_{~~\mu}  \\
\Psi^{II}_{~~\mu} & \bar{\Psi}^{I}_{~\mu}  & -i Im(\Theta_\mu) & 0  \\
\bar{\Psi}^{II}_{~~\mu} & \Psi^{I}_{~\mu}   & 0 & i Im(\Theta_\mu) \\
\end{bmatrix}\label{spin}
\eeq
where the complex valued functions $\Theta_\mu$, $\Psi^{I}_{~\mu}$ and $\Psi^{II}_{~~\mu}$ are defined by:
\begin{subequations}\label{DES_for_Lorentz}
\begin{align}
\Theta_\mu &\equiv  A^{-1}A_{,\mu} -  i \theta_{,\mu} - 2\bar{E}B_{,\mu}\\
\Psi^{I}_{~\mu} &\equiv A^{-1}e^{i\theta}(\bar{E}_{,\mu}-\bar{E}B_{,\mu}\bar{E})\\
\Psi^{II}_{~~\mu} &\equiv AB_{,\mu} e^{-i\theta}
\end{align}
\end{subequations}
in terms of the Lorentz parameter functions $\{A,\theta,B,E\}$ and where the comma indicates a partial derivative.
We can also represent the spin connection one-form $\bomega^a_{~b}$ in terms of the three one-forms $\Theta=\Theta_\mu\,dx^\mu$, $\Psi^I=\Psi^{I}_\mu\,dx^\mu$ and $\Psi^{II}=\Psi^{II}_{~~\mu}\,dx^\mu$.

\subsubsection{Spin Connection in an Orthonormal Gauge}

It is also possible to represent the most general spin connection in an {\bf{Orthonormal Gauge}} by applying a linear transformation
\beq
P^a_{~b} = \frac{1}{\sqrt{2}}\begin{bmatrix} 1 & 1 & 0 & 0 \\ 1 & -1 & 0 & 0 \\ 0 & 0 & 1 & 1 \\ 0 & 0 & -i & i \end{bmatrix}
\eeq
which will transform a {\bf{Complex Null Gauge}} to an {\bf{Orthonormal Gauge}}.  With this transformation the spin connection represented by equation \eqref{spin}
transforms to
\begin{equation}
\omega^{a}_{~b}=\begin{bmatrix}
0                       & Re(\Theta)       & Re(\Psi^{I}+\Psi^{II})\quad   &  -Im(\Psi^{I}-\Psi^{II})  \\
Re(\Theta)              & 0                & Re(\Psi^{I}-\Psi^{II})   &  -Im(\Psi^{I}+\Psi^{II})  \\
Re(\Psi^{I}+\Psi^{II})  & -Re(\Psi^{I}-\Psi^{II}) \quad & 0 &Im(\Theta)  \\
-Im(\Psi^{I}-\Psi^{II})\quad & Im(\Psi^{I}+\Psi^{II})  & -Im(\Theta) &0
\end{bmatrix}
\label{gen_spin_ortho}
\end{equation}
This expression provides a representation for a general Lorentz transformation in terms of the Lorentz parameter functions $\{A,\theta,B,E\}$ that can be used for orthonormal co-frames.

\subsection{TEGR and $F(T)$ Teleparallel gravity}\label{sec:BasicTEGR}

Let us consider TEGR and $F(T)$ theories.   Due to the differentiability of $M$, there exists a non-degenerate coframe
field $\bh^a$ which is defined on a subset $U$ of $M$. To achieve a notion of geometry (lengths and
angles) we also assume the existence of a symmetric metric field on $M$, with components $g_{ab}=g(\bh_a,\bh_b)$
with respect to this coframe field. In order to define a covariant differentiation process, we also assume the existence of a linear affine connection one form $\bomega^{a}_{~b}$.

With the assumption that the spin-connection be metric compatible, the spin-connection becomes anti-symmetric, $\bomega_{(ab)}=0$.  Due to the algebraic nature of this constraint, it can be implemented easily without loss of generality. The condition that the curvature vanishes imposes a second differential constraint on the spin-connection and implies that the most general solution to this equation is of the form:
\begin{equation}
\omega^a_{~b\mu} = (\Lambda^{-1})^{a}_{~b}\partial_\mu(\Lambda^b_{~c})\label{solution_omega}
\end{equation}
for some $\Lambda^a_{~b}\in SO^+(1,3)$.

The fundamental variables are the elements of the co-frame $\bh^a_{\phantom{a}}$.  The torsion associated with the coframe with the given spin-connection is:
\begin{equation}
T^a_{~\mu\nu}=\partial_\mu h^a_{~\nu}-\partial_\nu h^a_{~\mu}+\omega^a_{~b\mu}h^b_{~\nu}-\omega^a_{~b\nu}h^b_{~\mu}.
\end{equation}
To derive the field equations for teleparallel gravity, we consider a Lagrangian for $F(T)$ teleparallel gravity where the scalar quantity, $T$, is defined as
\begin{equation}
T =
\frac{1}{4} \; T^\rho_{\phantom{\rho}\mu\nu} \, T_\rho^{\phantom{\rho}\mu \nu} +
\frac{1}{2} \; T^\rho_{\phantom{\rho}\mu\nu} \, T^{\nu \mu}_{\phantom{\rho\rho}\rho} -
            T^\rho_{\phantom{\rho}\mu\rho} \, T^{\nu \mu}_{\phantom{\rho\rho}\nu}.
\label{TeleLagra}
\end{equation}
The scalar torsion $T$ can be written more compactly using the super-potential, $S^a_{~\mu \nu}$, expressed in terms of the torsion tensor as
\begin{equation}
S_a^{\phantom{a}\mu\nu}=\frac{1}{2}\left(T_a^{\phantom{a}\mu\nu}+T^{\nu\mu}_{\phantom{\nu\mu}a}-T^{\mu\nu}_{\phantom{\mu\nu}a}\right)-h_a^{\phantom{a}\nu}T^{\rho \mu}_{~~\rho} + h_a^{\phantom{a}\mu}T^{\rho \nu}_{~~\rho},
\end{equation}
so that:
\begin{equation}
T=\frac{1}{2}T^a_{\phantom{a}\mu\nu}S_a^{\phantom{a}\mu\nu}.
\end{equation}

The complete Lagrangian for $F(T)$ teleparallel gravity is
\begin{equation}
L= \frac{h}{2\kappa}F(T)+L_{Matt} \label{lagrangian}
\end{equation}
where $h$ is the determinant of the vielbein matrix $h^a_{~\mu}$ and $\kappa$ is the gravitational coupling constant, $\kappa=8\pi G/c^4$, where we have chosen units so that $c=1$.  We define the canonical energy momentum as follows
\begin{equation}
h\Theta_a^{\phantom{a}\mu}=\frac{\delta L_{Matt}}{\delta h^{\vphantom{\mu}a}_{\phantom{a}\mu}}.
\end{equation}
The variation of the Lagrangian \eqref{lagrangian} with respect to  $h^a_{\phantom{a}\mu}$, yields:
\begin{eqnarray}
\kappa \Theta_a^{\phantom{a}\mu}&=&
        h^{-1}F_T(T)\partial_v\left(hS_a^{\phantom{a}\mu\nu}\right)+F_{TT}(T)S_a^{\phantom{a}\mu\nu} \partial_v T \nonumber\\
        &&+\frac{1}{2}F(T)h_a^{\phantom{a}\mu} -F_T(T)T^b_{\phantom{a}a\nu}S_b^{\phantom{a}\mu\nu}-F_T(T)\omega^b_{\phantom{a}a\nu}S_b^{\phantom{a}\mu\nu}.\label{EQ2b}
 \end{eqnarray}
which is {\it Lorentz covariant}, and gives the field equations for determining the co-frame.

Since we have assumed the invariance of the field equations under $SO^+(1,3)$, the canonical energy momentum is symmetric $\Theta_{[ab]}=0$.
Further, it can be shown that the metrical energy momentum $T_{ab}$ now satisfies
\begin{equation}
T_{ab}\equiv\frac{1}{2}\frac{\delta L_{Matt}}{\delta g_{ab}}=\Theta_{(ab)}.
\end{equation}
Since $\Theta_{[ab]}=0$, from the anti-symmetric part of the  field equations  \eqref{EQ2b}, this condition then implies that
\begin{equation}
             0      = F_{TT}(T)S_{[ab]}^{\phantom{[ab]}\nu} \partial_v T.\label{asym2}
\end{equation}
In the case of TEGR, where $F(T) = T$, equation (\ref{asym2}) vanishes. It also vanishes for $T = T_0$, a constant.
For $F(T) \neq T$, it can be shown that the variation of the gravitational Lagrangian by the flat spin-connection is equivalent to the anti-symmetric part of the field equations \eqref{asym2}  \cite{Hohmann:2017duq,Golovnev:2017dox}. Thus if the field equations for the spin-connection are satisfied then the field equations for the coframe are guaranteed to be symmetric \cite{Hohmann:2018rwf}.


\subsection{Symmetry Algorithm}

We adapt the \emph{Symmetry Algorithm} introduced in \cite{Coley:2022aty} to study the Bianchi I simply transitive (i.e., no isotropies) spatially homogeneous geometries.  In this case the three vector generators of the affine frame symmetries ${\bf X}_i$ commute. The algorithm is:
\begin{enumerate}
\item  Choose canonical coordinates $x^i$ on the spatial hypersurface so that the Killing vectors ${\bf X}_i=\partial_{x^i}$. A set of left-invariant vector fields can be computed and the remaining freedom to do linear transformations allows one to express the metric components $g_{ij}$ in some desired form, such as, for example diagonal. As the orbits of the group of motions are non-null hyper-surfaces the fourth coordinate $x^0=t$ can be selected so that $g_{0i}=0$ and $g_{00}=-1$.  We keep the coordinates fixed thereafter. Essentially, the isometries present are used to choose coordinates adapted to the symmetries \cite{Stephani:2003tm}.

\item Construct a frame   ${\tilde{h}}^a$, such that
\be
g_{\mu\nu}=\eta_{ab}  {\tilde{h}}^{a}_{\phantom{a}\mu} {\tilde{h}}^{b}_{\phantom{a}\nu}, \quad {\text{where}}\quad \mathcal{L}_{{\bf X}}({\tilde{h}}^a) =0.
\ee

\item Since ${{\bf X}}$ is an affine frame symmetry, there exists a frame, and hence general Lorentz transformations
${\tilde{\Lambda}}^a_{\phantom{a}b}$ and ${\bar{\Lambda}}^a_{\phantom{a}b}$, such that
\be
{h}^a_{\phantom{a}\mu} ={\tilde{\Lambda}}^a_{\phantom{a}b}\tilde{h}^b_{\phantom{a}\mu},\ \
\omega^a_{\phantom{a}b\mu}={\bar{\Lambda}}^a_{\phantom{a}c} \partial_\mu ({\bar{\Lambda}}_b^{\phantom{a}c}),
\label{frame}
\ee
so that equations  (\ref{Intro:FS2}) imply that
\be
\mathcal{L}_{{\bf X}}({\tilde{\Lambda}}^a_{\phantom{a}b}) = 0: ~~
\mathcal{L}_{{\bf X}}({\bar{\Lambda}}^a_{\phantom{a}c} \partial_\mu ({\bar{\Lambda}}_b^{\phantom{a}c})) = 0.
\label{one} \ee

\item We can now affect another Lorentz transformation to simplify. Here we apply a Lorentz transformation
${\tilde{\Lambda}}^a_{\phantom{a}b}$ to set  the frame as the canonical frame ${\tilde{h}}^a$ and the spin connection is then of the general form (\ref{spin}).
\end{enumerate}


\section{Bianchi I geometries in $F(T)$ gravity}

\subsection{Bianch I geometry: the setup}

Bianchi type I geometries are simple anisotropic spatially homogeneous geometries. Spatial homogeneity occurs when a three dimensional group of symmetries acts simply transitively on space-like hyper-surfaces.  In the case in which the group of symmetries is of Bianchi type I, this group is commutative and the resulting spatial hyper-surface has zero curvature.

In any neighbourhood, for the Bianchi I geometry we can always find {\em{local}} (synchronous) coordinates $x^\mu=[t,x^j]=[t,x,y,z]$ where  $j \in\{1, 2, 3\}$.
We assume that the 3 space-like symmetry vectors (KVs) acting simply transitively on space-like hypersurfaces are denoted by $X_j  \equiv \frac{\partial}{\partial x^j}$.

Without loss of generality, the coordinate metric can be diagonalized, such that
\beq
g_{\mu\nu} = \rm{Diag}\left[ -1, R^2, P^2, Q^2 \right],
\label{metric} \eeq
where the metric functions $R, P, Q$ are  functions of $t$ alone.

Assuming a Null-frame gauge in which we have
\begin{equation} \eta_{ab}=\left[\begin{array}{cccc}
0 & -1 & 0 & 0 \\
-1 & 0 & 0 & 0 \\
0 & 0 & 0 & 1 \\
0 & 0 & 1 & 0
\end{array}\right] \end{equation}
we can select a co-frame basis 
\begin{equation} \label{coframe1}
h^a = \frac{1}{\sqrt{2}}\left[\begin{array}{c}
dt+ R(t) dx\\
dt- R(t) dx \\
P(t) dy+ iQ(t) dz \\
P(t) dy- iQ(t) dz
\end{array}\right]\end{equation}

%

To complete the geometrical setup, we need to determine the general form of the spin-connection one-form $\omega^a_{~b}$ that is invariant under the Bianchi I affine frame symmetry conditions. Since the affine symmetry vectors (Killing vectors) commute in the Bianchi type I geometry, we can use the results of \cite{Coley:2022aty} to obtain the general Lorentz transformation compatible with this symmetry assumption.
Briefly we need to solve
\begin{equation}
{\mathcal{L}}_{{\bf X}_j} \left[\omega^a_{~b\mu}\right]=0  \Rightarrow  {\mathcal{L}}_{{\bf X}_j} \left[(\Lambda^{-1}){}^a_{~c} \partial_\mu \Lambda^c_{~b} \right] =  0.\label{connection_condition}
\end{equation}
Obviously the components of $\omega^a_{~b\mu}=\omega^a_{~b\mu}(t)$ (i.e., are independent of $x^j$), but the Lorentz transformation $\Lambda^a_{~b}$ that produces the spin connection could still depend on $x^j$ albeit in a prescribed and restricted manner.

\subsection{Bianchi I Spin Connection}

Knowing that the components of the spin connection are functions of $t$ only permits one to partially determine the Lorentz parameter functions.  Equation \eqref{connection_condition} can have solutions in which the Lorentz parameter functions depend on $x^i$.   In general, one can solve the resulting differential equations \eqref{DES_for_Lorentz} for the Lorentz parameter functions. We define the following quantities
\begin{subequations}
\begin{eqnarray}
M_j&\equiv& L_j+iJ_j, \\
\lambda(x^j) &\equiv& M_j x^j = M_1x^1+M_2x^2+M_3 x^3,\\
\lambda_{Re}(x^j) &\equiv& \Re(M_jx^j)=L_jx^j =  L_1x+L_2y+L_3 z,\\
\lambda_{Im}(x^j) &\equiv& \Im(M_jx^j) =J_jx^j = J_1x+J_2y+J_3 z,
\end{eqnarray}
\end{subequations}
where $j\in \{1,2,3\}$ and $M_j$ is a complex valued constant ($L_j$ and $J_j$ are real valued constants). Here we expand upon the solution determined in \cite{Coley:2022aty} where $J_j=0$ was used to construct a general solution.
A general solution to \eqref{DES_for_Lorentz} in the case of Bianchi I symmetry is
\begin{subequations}\label{General Solution}
\begin{align}
A(t,x^j)e^{-i\theta(t,x^j)}&=\alpha(t)e^{-i\vartheta(t)}e^{\lambda(x^j)},\\
\bar{E}(t,x^j)&=\bar{\eta}(t)e^{\lambda(x^j)},\\
B(t,x^j) &= \beta(t)e^{-\lambda(x^j)},
\end{align}
\end{subequations}
where $\alpha(t)$ and $\vartheta(t)$ are real valued functions, $\bar{\eta}(t)$ and $\beta(t)$ are complex valued functions.
Alternatively we can express
\begin{subequations}\label{General Solutionpart2}
\begin{align}
A(t,x^j)&=\alpha(t)e^{\lambda_{Re}(x^j)},\\
\theta(t,x^j) &= \vartheta(t)-\lambda_{Im}(x^j).
\end{align}
\end{subequations}
The general solution to \eqref{DES_for_Lorentz} given by \eqref{General Solution} contains six real valued constants of integration contained in $M_j$ and six arbitrary functions, $\alpha(t)$, $\vartheta(t)$ and the real and imaginary parts of $\bar{\eta}(t)$ and $\beta(t)$. From here on we no longer explicitly show the functional dependence on the coordinates $t$ and $x^j$.

While the Lorentz parameter functions depend on the coordinates $t$ and $x^j$, the components of the spin connection $\Theta_\mu$, $\Psi^I_{~\mu}$ and $\Psi^{II}_{~~\mu}$ are independent of $x^j$ as expected and reduce to
\begin{subequations}\label{Special_Spin_connection2}
  \begin{align}
  \Theta_0 &= \frac{\dot{\alpha}}{\alpha}-i\dot\vartheta-2\bar{\eta}\dot\beta,
                   &\Theta_j &= M_j(1+2\bar{\eta}\beta),\\
  \Psi^I_{~0} &= \left(\dot{\bar{\eta}}-\bar{\eta}^2\dot\beta\right)\alpha^{-1}e^{i\vartheta}, \qquad
                   &\Psi^I_{~j} &=M_j\left(\bar{\eta}+\bar{\eta}^2\beta\right)\alpha^{-1}e^{i\vartheta}, \\
  \Psi^{II}_{~~0} &= \alpha e^{-i\vartheta}\dot\beta, &\Psi^{II}_{~~j}&=-M_j\alpha e^{-i\vartheta}\beta,
  \end{align}
\end{subequations}
where an over-dot corresponds to the time derivative.

For completeness we present the corresponding Lorentz transformation \eqref{Special_Spin_connection2}
\begin{equation} \label{general_Lorentz}
\Lambda_{~c}^{b}= \begin{bmatrix}
\alpha e^{\lambda_{Re}} &
\alpha^{-1}\eta\bar{\eta} e^{\lambda_{Re}} &
\bar{\eta}e^{i\vartheta} e^{\lambda_{Re}} &
\eta e^{-i\vartheta} e^{\lambda_{Re}}\\
\alpha\beta\bar{\beta} e^{-\lambda_{Re}} &
\alpha^{-1}(1+\beta\bar{\beta}\eta\bar{\eta}+\bar{\beta}\eta+\beta\bar{\eta}) e^{-\lambda_{Re}} & (\beta\bar{\beta}\bar{\eta}+\bar{\beta})e^{i\vartheta} e^{-\lambda_{Re}} &
(\beta\bar{\beta}\eta+\beta)e^{-i\vartheta} e^{-\lambda_{Re}}  \\
\alpha\beta e^{-i\lambda_{Im}} &
\alpha^{-1}(\beta\eta\bar{\eta}+\eta) e^{-i\lambda_{Im}} &
(1+\beta\bar{\eta})e^{i\vartheta} e^{-i\lambda_{Im}} &
\beta\eta e^{-i\vartheta}e^{-i\lambda_{Im}}  \\
\alpha\bar{\beta} e^{i\lambda_{Im}}  &
\alpha^{-1}(\bar{\beta}\eta\bar{\eta}+\bar{\eta})e^{i\lambda_{Im}} &
\bar{\beta}\bar{\eta}e^{i\vartheta}e^{i\lambda_{Im}} &
(1+\bar{\beta}\eta)e^{-i\vartheta}e^{i\lambda_{Im}} \\
\end{bmatrix}
\end{equation}
which contains six arbitrary real valued functions and six arbitrary constants. This general solution for the Lorentz transformation represents a situation in which the Lie derivative of the Lorentz matrix transformation is not zero, but the Lie derivative of the spin connection is zero,
\begin{equation}
{\mathcal{L}}_{{\bf X}_j} \left[\Lambda^a_{~b}\right]\not=0,\qquad {\mathcal{L}}_{{\bf X}_j} \left[\Lambda^a_{~c}d\,(\Lambda^{-1})^c_{~b}\right]=0.
\end{equation}
This is because of the dependence on $x^j$ present in $\lambda_{Re}$ and $\lambda_{Im}$ in equation \eqref{general_Lorentz}.

\subsection{The Antisymmetric part of the $F(T)$ field equations}

Now that we know the nature of the most general spin connection suitable for Bianchi I geometries from equations \eqref{spin} and \eqref{Special_Spin_connection2}, we wish to ascertain what constraints, if any, the antisymmetric part of the field equations for $F(T)$ tele-parallel gravity place on the functions $\alpha(t),\vartheta(t),\beta(t),\eta(t)$ and on the constants $M_j$ in the spin connection. We first note that $T$ is a function of $t$ only, as expected from the symmetry assumption.  We recall that the equations of motion for the spin connection, which result from varying the action with respect to the spin connection in $F(T)$ teleparallel gravity, is equivalent to the antisymmetric part of the field equations resulting from varying with respect to the frame \cite{Golovnev:2017dox}.  We also recall, that if we have a diagonal Bianchi I co-frame and the spin connection is identically zero then the antisymmetric part of the field equations are identically satisfied.  Since we are currently looking at the most general spin connection possible, this may not be true here. A number of different classes of solutions to the anti-symmetric part of the field equations \eqref{asym2} are possible.

\subsubsection{TEGR}
Equations \eqref{asym2} are identically satisfied if $F_{TT}(T)=0$ and thus $F(T)$ is linear in $T$, but this is TEGR which we will not consider further here.

\subsubsection{Constant $T$}
A second solution arises if $T$ is a constant. The torsion scalar $T$ is a non-linear function of the two real and two complex functions $\alpha(t),\vartheta(t),\beta(t),\eta(t)$ and their derivatives. Loosely speaking in this case, we have one first order differential equation containing six arbitrary functions and six constants.  There will exist multiple solutions which will yield a constant value for $T$ including the value zero.   If $T$ is indeed made to be constant via a choice of $\alpha(t),\vartheta(t),\beta(t)$, and $\eta(t)$, then the field equations are equivalent to a re-scaled version of TEGR (which looks like GR with a cosmological constant and a re-scaled coupling constant) \cite{Krssak:2018ywd}. This possibility is not surprising and is not considered here.

\subsubsection{Non-Constant T}
Of course, more interesting and non-GR type results occur when $T$ is not a constant and $F(T)$ is nonlinear.  If $T$ is not constant and $F(T)$ is nonlinear then we have that the non-trivial antisymmetric part of the field equations require
\begin{equation}
S_{[ab]}^{\phantom{[ab]}t} = 0.
\end{equation}
In general, the components of ${S_{[ab]}}^t$ in terms of $\Theta$, $\Psi^I$ and $\Psi^{II}$ are:
\begin{widetext}
\begin{eqnarray*}
{S_{[01]}}^{t}&=&
\frac{1}{4P}\,\left(\Psi^{I}_{~2}+\overline{\Psi}^{I}_{~2} - \Psi^{II}_{~~2}-\overline{\Psi}^{II}_{~~2}\right)
+\frac{i}{4Q}\,\left(\Psi^{I}_{~3}-\overline{\Psi}^{I}_{~3}+ \Psi^{II}_{~~3}-\overline{\Psi}^{II}_{~~3}\right),
\\
{S_{[03]}}^{t}&=& \frac{1}{4}\frac{\Theta_2}{P}-\frac{1}{2}\frac{\Psi^{II}_{~~1}}{R}
                 -\frac{i}{4}\frac{\Theta_3}{Q},
\\
{S_{[12]}}^{t}&=&-\frac{1}{4}\frac{\Theta_2}{P}+\frac{1}{2}\frac{\Psi^{I}_{~1}}{R}
                 +\frac{i}{4}\frac{\Theta_3}{Q},
\\
{S_{[23]}}^{t}&=&
\frac{1}{4P}\,\left(\Psi^{I}_{~2}-\overline{\Psi}^{I}_{~2} - \Psi^{II}_{~~2}+\overline{\Psi}^{II}_{~~2}\right)
+\frac{i}{4Q}\,\left(\Psi^{I}_{~3}+\overline{\Psi}^{I}_{~3}+ \Psi^{II}_{~~3}+\overline{\Psi}^{II}_{~~3}\right),
\end{eqnarray*}
\end{widetext}
where we note that the complex conjugates of ${S_{[03]}}^{t}$ and ${S_{[12]}}^{t}$ yield the two remaining components ${S_{[02]}}^{t}$ and ${S_{[12]}}^{t}$ respectively. Solving  ${S_{[ab]}}^t=0$ yields the following three linear relationships for some of the components of the complex valued functions $\Theta_{\mu}$, $\Psi^{I}_{~\mu}$ and $\Psi^{II}_{~~\mu}$.
\begin{subequations}
\begin{align}
\frac{2 \Psi^{I}_{~1}}{R}&=\frac{\Theta_2}{P}-i \frac{\Theta_3}{Q},\\
\frac{2 \Psi^{II}_{~1}}{R}&=\frac{\Theta_2}{P}+i \frac{\Theta_3}{Q},\\
\frac{  -\Psi^{I}_{~2}+\Psi^{II}_{~~2}}{P} &= i\frac{  \Psi^{I}_{~3}+\Psi^{II}_{~~3}}{Q}.
\end{align}
\end{subequations}
Substituting the solution for the spin connection \eqref{Special_Spin_connection2} into the conditions for ${S_{[ab]}}^{t}=0$ we obtain constraints on the arbitrary functions $\alpha(t)$, $\theta(t)$, $\beta(t)$, and $\bar{\eta}(t)$. These conditions yield the following
\begin{subequations}\label{constraints}
\begin{eqnarray}
  \frac{2M_1}{R}\left(\bar{\eta}+\bar{\eta}^2\beta\right)\alpha^{-1}e^{i\vartheta}
      -\left(\frac{M_2}{P}-i\frac{M_3}{Q}\right)\left(1+2\bar{\eta}\beta\right)&=&0,\\
  \frac{2M_1}{R}\beta\alpha e^{-i\vartheta}+\left(\frac{M_2}{P}+i\frac{M_3}{Q}\right)\left(1+2\bar{\eta}\beta\right)&=&0,\\
  \left(\frac{M_2}{P}+i\frac{M_3}{Q}\right)\left(\bar{\eta}+\bar{\eta}^2\beta\right)\alpha^{-1}e^{i\vartheta}+\left(\frac{M_2}{P}-i\frac{M_3}{Q}\right)\beta\alpha e^{-i\vartheta} &=& 0.
\end{eqnarray}
\end{subequations}
Explicit solutions to \eqref{constraints} can be classified according to the values of the integration constants $M_j$.

\paragraph{General Solution: $M_1=M_2=M_3=0$.} \label{gen_sol}

If $M_1=M_2=M_3=0$ then the antisymmetric part of the field equations represented by equations \eqref{constraints} are identically satisfied. In this case since we still have six arbitrary functions of $t$; we obtain a {\bf general solution},
    \begin{equation}\label{general_solution1}
    A(t,x^j)=\alpha(t),\qquad \theta(t,x^j)=\vartheta(t),\qquad B(t,x^j)=\beta(t),\qquad \bar{E}(t,x^j)=\bar{\eta}(t).
    \end{equation}
The corresponding Lorentz transformation associated with this solution is independent of the spatial coordinates $x^j$, that is,  $ {\mathcal{L}}_{{\bf X}_j} \left[\Lambda^a_{~b}\right]=0$ for $j\in\{1,2,3\}$.  We could apply this time dependent Lorentz transformation to obtain a zero spin connection and a proper co-frame $\bh'^a$. The resulting proper co-frame will satisfy $ {\mathcal{L}}_{{\bf X}_j} \left[\bh'^a\right]=0$ for $j\in\{1,2,3\}$.   However, this proper co-frame $\bh'^a$ is no longer of the simple form \eqref{coframe1} and will contain nine arbitrary functions of $t$.

\paragraph{Special Solution: $M_1\not = 0$, $M_2=M_3=0$.}\label{spec_sol}

If $M_1\not = 0$ and $M_2=M_3=0$ then we obtain a {\bf special solution}.  We observe that $M_2=M_3=0$ if and only if $\bar{\eta}=0$ and $\beta=0$. This means  the resulting solution contains only two arbitrary real-valued functions and two arbitrary real-valued constants,
    \begin{equation}
    A(t,x^j)=\alpha(t)e^{L_1x}, \qquad \theta(t,x^j)=\vartheta(t)-J_1x,\qquad B(t,x^j)=0, \qquad \bar{E}(t,x^j)=0
    \end{equation}
Here, the corresponding Lorentz transformation associated with this solution is dependent upon the spatial coordinate $x$, that is,  $ {\mathcal{L}}_{{\bf X}_1} \left[\Lambda^a_{~b}\right]\not=0$.
Therefore, if we apply this Lorentz transformation to obtain a zero spin connection and a proper co-frame $\bh'^a$, then $ {\mathcal{L}}_{{\bf X}_j} \left[\bh'^a\right]\not = 0$ for $j\in\{1,2,3\}$.  However the underlying geometry remains Bianchi type I, that is, the 3-dimensional zero curvature spatial hyper-surface is still present, it is only that the proper frame chosen in this case is not fully aligned with the symmetry of the situation. 

\paragraph{Special Solution: $M_2\not = 0$ or $M_3\not = 0$}\label{spec_sol2}

All other possibilities to solve \eqref{constraints} will have either $M_2\not = 0$ or $M_3\not = 0$.  If $M_2\not = 0$ or $M_3\not = 0$ then the resulting solution is special in that it contains two arbitrary real-valued functions $\alpha(t),\vartheta(t)$ and up to 3 arbitrary complex-valued constants $M_j=L_j+iJ_j$. The solution to \eqref{constraints} is
\begin{subequations}
\begin{eqnarray}
A(t,x^j)          &=&\alpha(t)e^{L_1x+L_2y+L_3z},\\
\theta(t,x^j)     &=&\vartheta(t)-J_1x-J_2y-J_3z,\\
B(t,x^j)             &=& \mp \frac{1}{2\widehat{M}(t)}\left(\frac{M_2}{P(t)}+i\frac{M_3}{Q(t)}\right)\alpha(t)^{-1} e^{i\vartheta(t)},\label{sol_beta}\\
\bar E(t,x^j)         &=& \left(-\frac{M_1}{R(t)}\pm\widehat{M}(t)\right)\left(\frac{M_2}{P(t)}+i\frac{M_3}{Q(t)}\right)^{-1}\alpha(t) e^{-i\vartheta(t)}\label{sol_eta}
\end{eqnarray}
\end{subequations}
where we define the complex valued function
$$\widehat{M}(t)\equiv\sqrt{\left(\frac{M_1}{R(t)}\right)^2+\left(\frac{M_2}{P(t)}\right)^2+\left(\frac{M_3}{Q(t)}\right)^2}.$$
We note that $B(t,x^j)$ and $E(t,x^j)$ are functions of $t$ only.  Spatial dependence arises only in the boost and spin parameters $A$ and $\theta$.

Similar to the previous special solution, the corresponding Lorentz transformation associated with this solution is again dependent of the spatial coordinates $x^j$.  Therefore, if we apply this Lorentz transformation to obtain a zero spin connection and a proper co-frame, then the resulting proper co-frame will not satisfy the Bianchi I symmetry requirement, although the underlying geometry is Bianchi type I.

\subsection{The Symmetric Part of the $F(T)$ Field Equations for non-constant $T$}

Having an ansatz for the spin connection in the case of Bianchi I symmetries, as determined by the Lorentz parameter functions above, and given the Bianchi I co-frame \eqref{coframe1}, we have shown that the antisymmetric part of the $F(T)$ field equations are now satisfied. That is, we have determined the most general class of ``good tetrads'' \cite{Tamanini_Boehmer2012} to be used in Bianchi I teleparallel geometries in which the torsion scalar is not constant.

If the torsion scalar is a constant then the symmetric part of the teleparallel field equations reduce to the equivalent of GR with a cosmological constant and a re-scaled coupling constant.  Here we are particularly interested in teleparallel geometries in which the torsion scalar is not constant.  We shall investigate the differences in the symmetric part of the $F(T)$ teleparallel field equations for the first two solutions \ref{gen_sol} and \ref{spec_sol}. The challenges that can occur in choosing the co-frame spin connection pair can be illustrated without looking at the more complicated special solution presented in the third scenario \ref{spec_sol2}. We will write down the symmetric part of the field equations for the general solution $M_1=M_2=M_3=0$ and the special solution $M_1\not = 0$ and $M_2=M_3=0$ for illustrative purposes.

Most studies of spatially homogeneous solutions, particularly within the cosmological context, assume a perfect fluid source, which is a well defined and physical matter field. Here, we shall assume a comoving perfect fluid with energy density $\rho$ and isotropic pressure $p$. Anisotropic fluid sources are more problematic. Either unrealistic equations of state must be assumed (and exact solutions are very unlikely) or the field equations simply lead to a definition of the matter source and the resulting geometry is neither an exact solution nor physically interesting.

\subsubsection{General Solution: $M_1=M_2=M_3=0$}

In this case  the torsion scalar becomes
\begin{equation}
T = 2\frac{\dot{R}\dot{Q}}{RQ} +  2\frac{\dot{R}\dot{P}}{RP} + 2\frac{\dot{P}\dot{Q}}{PQ}.
\end{equation}
The symmetric part of the field equations yield the following set of linearly independent equations
\begin{subequations}\label{FE-1}
    \begin{align}
    \kappa\rho &=-\frac{1}{2}\big(F(T)-TF'(T)\big)+F'(T)\left(\frac{\dot{P}}{P}\frac{\dot{Q}}{Q}+\frac{\dot{Q}}{Q}\frac{\dot{R}}{R} +\frac{\dot{P}}{P}\frac{\dot{R}}{R}\right) ,  \\
    \kappa p&= \frac{1}{2}\big(F(T)-TF'(T)\big)-F'(T)\left( \frac{\ddot{P}}{P} +\frac{\ddot{Q}}{Q}
           +\frac{\dot{Q}}{Q}\frac{\dot{P}}{P}\right)
    - F''(T)\dot{T}\left(\frac{\dot{P}}{P}+\frac{\dot{Q}}{Q} \right),\\
    \kappa p&=  \frac{1}{2}\big(F(T)-TF'(T)\big)-F'(T)\left( \frac{\ddot{P}}{P} +\frac{\ddot{R}}{R}
           +\frac{\dot{P}}{P}\frac{\dot{R}}{R}\right)
    - F''(T)\dot{T}\left(\frac{\dot{P}}{P}+\frac{\dot{R}}{R} \right),\\
    \kappa p&=  \frac{1}{2}\big(F(T)-TF'(T)\big)-F'(T)\left( \frac{\ddot{Q}}{Q} +\frac{\ddot{R}}{R}
           +\frac{\dot{Q}}{Q}\frac{\dot{R}}{R} \right)
    - F''(T)\dot{T}\left(\frac{\dot{Q}}{Q}+\frac{\dot{R}}{R} \right),
    \end{align}
\end{subequations}
which is obviously independent of the Lorentz parameter functions. We will obtain a ``6'' Lorentz parameter set of solutions to the field equations.  The energy momentum conservation equation has the form
\begin{equation}
\dot\rho = -\left( \frac{\dot{R}}{R}+\frac{\dot{P}}{P}+\frac{\dot{Q}}{Q}\right)(\rho+p). \label{EMconservation}
\end{equation}

Rearranging the last three equations of \eqref{FE-1} we can obtain two linearly independent homogeneous expressions of the form
\begin{subequations}\label{eq43}
\begin{align}
0&=  -F'(T)\left( \frac{\ddot{Q}}{Q} -\frac{\ddot{R}}{R} +\frac{\dot{P}}{P}\left(\frac{\dot{Q}}{Q}-\frac{\dot{R}}{R}\right) \right)
    - F''(T)\dot{T}\left(\frac{\dot{Q}}{Q}-\frac{\dot{R}}{R} \right),\\
0&=  -F'(T)\left( \frac{\ddot{P}}{P} -\frac{\ddot{Q}}{Q}
           +\frac{\dot{R}}{R}\left(\frac{\dot{P}}{P}-\frac{\dot{Q}}{Q}\right)\right)
    - F''(T)\dot{T}\left(\frac{\dot{P}}{P}-\frac{\dot{Q}}{Q} \right).
\end{align}
\end{subequations}
In the isotropic special case in which $P \propto Q$ and $P\propto R$ equations \eqref{eq43} are identically satisfied. If only one of $P \propto Q$ or $P\propto R$ or $Q\propto R$ then equation \eqref{eq43} reduces to only one differential equation and is a locally rotationally symmetric (LRS) subcase of the general case.  In the general case equations \eqref{eq43} can be integrated to yield
\begin{subequations}\label{intFE}
  \begin{align}
  F'(T)(PQR)\left(\frac{\dot{Q}}{Q}-\frac{\dot{R}}{R}\right) &= C_1, \\
  F'(T)(PQR)\left(\frac{\dot{P}}{P}-\frac{\dot{Q}}{Q}\right) &= C_2,
  \end{align}
\end{subequations}
where $C_1$ and $C_2$ are arbitrary constants.   If $C_1=C_2=0$ then $P \propto Q$ and $P\propto R$ and we obtain the $k=0$ TRW geometry  after a simple rescaling of the coordinates \cite{Coley:2022qug}.  If only one of the constants is zero, or if $C_1=C_2$ then we reduce to one of the LRS special cases mentioned above. However, as long as one or both of the arbitrary constants are not zero, we obtain
\begin{equation}
\left(\frac{P}{R}\right)^{C_2}=\left(\frac{Q}{R}\right)^{C_1},
\end{equation}
regardless of the assumptions on the perfect fluid matter source.

\subsubsection{Special Solution: $M_1\not=0$, $M_2=M_3=0$ }
Clearly in this case we have $B_{,\mu}=0$ and $\bar{E}_{,\mu}=0$.  The torsion scalar and the symmetric part of the field equations in this case are similar to those in the general case with slight modifications. The torsion scalar becomes
\begin{equation}
T = 2\frac{\dot{R}\dot{Q}}{RQ} +  2\frac{\dot{R}\dot{P}}{RP} + 2\frac{\dot{P}\dot{Q}}{PQ} -  2\frac{L_1}{R}\left(\frac{\dot{P}}{P}+\frac{\dot{Q}}{Q}\right).
\end{equation}
 
The symmetric part of the field equations becomes
\begin{subequations}\label{FE-2}
    \begin{align}
    \kappa\rho &=-\frac{1}{2}\big(F(T)-TF'(T)\big)+F'(T)\left(\frac{\dot{P}}{P}\frac{\dot{Q}}{Q}+\frac{\dot{Q}}{Q}\frac{\dot{R}}{R} +\frac{\dot{P}}{P}\frac{\dot{R}}{R}\right),\label{FE-2a}\\
    \kappa p&= \frac{1}{2}\big(F(T)-TF'(T)\big)-F'(T)\left( \frac{\ddot{P}}{P} +\frac{\ddot{Q}}{Q}
           +\frac{\dot{Q}}{Q}\frac{\dot{P}}{P}\right)
    - F''(T)\dot{T}\left(\frac{\dot{P}}{P}+\frac{\dot{Q}}{Q} \right),\label{FE-2b}\\
     \kappa p&=  \frac{1}{2}\big(F(T)-TF'(T)\big)-F'(T)\left( \frac{\ddot{P}}{P} +\frac{\ddot{R}}{R}
           +\frac{\dot{P}}{P}\frac{\dot{R}}{R}\right)
    - F''(T)\dot{T}\left(\frac{\dot{P}}{P}+\frac{\dot{R}}{R} - \frac{L_1}{R} \right),\label{FE-2c}\\
    \kappa p&=  \frac{1}{2}\big(F(T)-TF'(T)\big)-F'(T)\left( \frac{\ddot{Q}}{Q} +\frac{\ddot{R}}{R}
           +\frac{\dot{Q}}{Q}\frac{\dot{R}}{R} \right)
           - F''(T)\dot{T}\left(\frac{\dot{Q}}{Q}+\frac{\dot{R}}{R}  - \frac{L_1}{R} \right).\label{FE-2d}
    \end{align}
\end{subequations}
Note how the torsion scalar and field equations are independent of the rotation parameter $J_1$ but do contain the boost parameter $L_1$. The rotation parameter $J_1$ affects the subspace spanned by $y$ and $z$ and therefore because of the symmetry assumptions, there should be no effect.  However, the boost parameter $L_1$ produces a transformation in the subspace spanned by $t$ and $x$, thereby transforming one of the basis vectors that were originally tangent to the surfaces of homogeneity to now be non-tangential. The energy momentum conservation equation is the same as (\ref{EMconservation}).

Similarly to the general case, by subtracting \eqref{FE-2d} from  \eqref{FE-2c} we can obtain expression of the form
\begin{align}
0&=  -F'(T)\left( \frac{\ddot{P}}{P} -\frac{\ddot{Q}}{Q}
           +\frac{\dot{R}}{R}\left(\frac{\dot{P}}{P}-\frac{\dot{Q}}{Q}\right)\right)
    - F''(T)\dot{T}\left(\frac{\dot{P}}{P}-\frac{\dot{Q}}{Q} \right).\label{temp123}
\end{align}
If $P\propto Q$ this expression is identically satisfied.  In general equation \eqref{temp123} can be integrated to yield
  \begin{align}
  F'(T)(PQR)\left(\frac{\dot{P}}{P}-\frac{\dot{Q}}{Q}\right) &= C_2, \label{intFE2}
  \end{align}
where $C_2$ is an arbitrary constant.   If $C_2=0$ then we have the special case $P\propto Q$. Further, equation \eqref{intFE2} is true independent of the assumptions on the perfect fluid.


\section{Bianchi type I Power-law Solutions}

The Bianchi type I geometry is the anisotropic generalization of the $k=0$ TRW geometry. It can be said that Bianchi I class of geometries in GR yield some of the most important classes of anisotropic solutions to the GR field equations.  Indeed, it is hypothesized that GR approaches an infinite cycle of different vacuum Bianchi type I solutions as one goes back in time \cite{ellis_uggla_wainwright_1997,hobill_1997,Coley:2003mj} resulting in the Belinski, Khalalatnikov, Lifshitz (BKL) dynamics.

In GR these vacuum Bianchi type I solutions are called the Kasner solutions and play an integral role in past behaviour of both anisotropic and inhomogeneous cosmological models.  In GR the frame (metric) functions have the form
\beq
P(t) = P_0\,t^p, \qquad &Q(t) = Q_0\,t^q,\qquad &R(t) = R_0\,t^r ,\label{kasner}
\eeq
in which the vacuum GR field equations impose the following two Kasner conditions
\beq
p+q+r=1, \qquad \mbox{and} \qquad p^2+q^2+r^2=1.\label{Kasner_relations}
\eeq
The constants $P_0$, $Q_0$ and $R_0$ can all be set to 1 without loss of generality via a coordinate transformation.
We will determine the conditions in which power-law solutions exist in $F(T)$ teleparallel gravity.

\subsection{Vacuum: General Case $M_1=M_2=M_3=0$}

The field equations \eqref{FE-1} in vacuum yield expressions in $F(T)$, $F'(T)$ and $F''(T)$. Assuming $F(T)\not = T$, the first field equation can be expressed as
\begin{equation}
0=-\frac{1}{2}F(T)+TF'(T)\label{vacuum_firedmann}
\end{equation}
where we immediately conclude that
\begin{equation}
F(T)=F_0\sqrt{T}.\label{vacuum_sol}
\end{equation}
We note that the differential equation \eqref{vacuum_firedmann} is singular when $T=0$, therefore a second solution to \eqref{vacuum_firedmann} is to require
\begin{equation}
T=0 \text{\ and\ } \lim_{T\to 0}F(T) =0, \label{singular_solution}
\end{equation}
as was presented in \cite{Paliathanasis:2016vsw,Paliathanasis:2017htk}. In this singular case, the power-law solution satisfies the same Kasner relations \eqref{Kasner_relations} as in GR and the torsion scalar is constant. Therefore, we will not consider it further here.

In the general case, with the power-law ansatz as given in equation \eqref{kasner}, where the spin connection is determined by the Lorentz parameter functions \eqref{general_solution1},  the three remaining field equations after solving \eqref{vacuum_firedmann} yield
\begin{subequations}\label{Vacuum_BI}
  \begin{align}
  p^2+q^2 &= r(p+q),\\
  p^2+r^2 &= q(p+r),\\
  q^2+r^2 &= p(q+r).
  \end{align}
\end{subequations}
One solution is to have $p=q=r=0$, but this yields the constant value of zero for the torsion scalar. The only non-trivial torsion scalar permitted by the field equations \eqref{Vacuum_BI} is the isotropic vacuum solution where $p=q=r \not = 0$ in which case the torsion scalar is $T=6p^2t^{-2}$. This solution is clearly the isotropic special case of equation \eqref{intFE}.

Therefore, anisotropic power-law solutions to the $F(T)\not = T$ teleparallel gravity field equation do not exist. Any Kasner-like solutions in this general case necessarily require consideration of the singular solution \eqref{singular_solution} which essentially reduces to GR with a re-scaled gravitational coupling.  The only power-law solution that exists in the general case having a non-trivial torsion scalar is the isotropic $k=0$ TRW geometry.

\subsection{Vacuum: Special Solution  $M_1\not=0$, $M_2=M_3=0$}

The vacuum field equations yield expressions for $F(T)$, $F'(T)$ and $F''(T)$, but unlike in the general case, one cannot immediately integrate the first field equation to find the closed form of $F(T)$ as a function of $T$. However, in vacuum we can use equations \eqref{intFE2} and \eqref{FE-2a}  to determine an expression for $F$ as a function of $t$.  We again note that $L_1\not=0$ else we return to a sub-case of the general case considered above. Due to the existence (or not) of the first integral \eqref{intFE2}, this case breaks into two subcases. We are interested in power-law solutions in which the torsion scalar is not constant.

\subsubsection{$P$ not proportional to $Q$}
Substituting the power-law relations in equation \eqref{kasner} into the field equation \eqref{FE-2a} and the first integral \eqref{intFE2}, we obtain for $P\not \propto Q$ (or $p\not = q$ or $C_2\not = 0$)
\begin{subequations}\label{temp}
  \begin{align}
  F'(T) &= \frac{C_2}{p-q} t^{-(p+q+r-1)}, \label{F-prime1}\\
  \frac{1}{2}\big(F-TF'(T)\big) &= \frac{C_2}{p-q}(pq+pr+qr) t^{-(p+q+r+1)},
  \end{align}
\end{subequations}
which also allows us to compute
\begin{equation}\label{temp2}
F''(T)\dot T = \frac{\partial}{\partial t}\big(F'(T)\big) = -\frac{C_2}{p-q}(p+q+r-1)t^{-(p+q+r)}.
\end{equation}
Substituting equations \eqref{temp} and \eqref{temp2} in the remaining field equations \eqref{FE-2} we find that
\begin{subequations}
    \begin{align}
    \frac{C_2}{p-q}t^{-(p+q+r+1)}\big(2pq+2pr+2pq\big) &=0,\\
    \frac{C_2}{p-q}t^{-(p+q+r+1)}\big(2pq+2pr+2pq-L_1(p+q+r-1)t^{(1-r)}\big) &=0,\\
    \frac{C_2}{p-q}t^{-(p+q+r+1)}\big(2pq+2pr+2pq-L_1(p+q+r-1)t^{(1-r)}\big) &=0.
    \end{align}
\end{subequations}
Since $L_1\not = 0$, the solution to the field equations is
\begin{equation}
p+q+r=1,\qquad pq+pr+qr=0,
\end{equation}
which is equivalent to the Kasner relations \eqref{Kasner_relations}.  The torsion scalar becomes
\beq
T=2L_1(r-1)t^{-(r+1)},
\eeq
but for this solution, equation \eqref{F-prime1} implies that $F(T)$ is linear in $T$ and we are back to a re-scaled TEGR.

\subsubsection{$P \propto Q$}

If $P \propto Q$  then we no longer have the first integral \eqref{intFE2}, but we are able to make significant progress nonetheless. Since $p=q$, equation \eqref{FE-2a} can be rewritten as
\begin{equation}
 \frac{1}{2}\big(F-TF'(T)\big) = F'(T)(p^2+2pr)t^{-2}, \label{F-prime2}\\
\end{equation}
Using this expression and the power-law relations \eqref{kasner} we can express the other equations in \eqref{FE-2} as
\begin{subequations}
  \begin{align}
  2p\left[F'(T)(r+1-p)t^{-2}-F''(T)\dot{T} t^{-1}\right] &=0 \label{thisone}\\
  F'(T)(pr+p+r-r^2)t^{-2}-F''(T)\dot{T}\big((p+r)t^{-1}-L_1t^{-r}\big) &=0
  \end{align}
\end{subequations}
Here we see two cases
\begin{enumerate}
\item $p=q=0$:  In this case $r=1$ or $r=0$ and the torsion scalar reduces to a constant trivial torsion scalar $T=0$.
\item $p=q=\frac{2}{3}$, $r=-\frac{1}{3}$.  The torsion scalar becomes $T=-\frac{8}{3}L_1t^{-2/3}$.  However, when this solution is substituted back into equation \eqref{thisone} we find that $F(T)$ is linear in $T$ and we are back to TEGR.
\item $p=q$, $r=1$, $L_1=\frac{(p-1)(p+2)}{p-2}$ and $p\not=2$.  Here we obtain $F(T)=F_0 T^{p/2}$ with $-2< p<2$. This is an anisotropic vacuum power-law solution with a non-trivial torsion scalar
\beq
T=\frac{-2p^2(p+2)}{p-2}t^{-2}.
\eeq
Here the line element is of the form
\beq ds^2 = - dt^2 + t^2 dx^2 + t^{2p} dy^2 + t^{2p} dz^2. \label{spec_vac}
\eeq
For the spin connection in the null frame, assuming all the arbitrary functions are zero and only keeping the non-zero $L_1$, we have $\omega^1_{~1x}=-\omega^{2}_{~2x}=L_1$ (all other components can be made to be zero without loosing the salient feature here). The Lorentz transformation that yields this connection is
\begin{equation}
\Lambda^a_{~b}=
\begin{bmatrix}
  e^{L_1x} & 0 & 0 & 0  \\
  0 & e^{-L_1x} & 0 & 0  \\
  0 & 0 & 1 & 0  \\
  0 & 0 & 0 & 1
\end{bmatrix}.
\end{equation}
\end{enumerate}

\subsection{Non-vacuum Perfect Fluid Field Equations}

We can look for non-vacuum power-law solutions in $F(T)$ teleparallel gravity when the geometry is Bianchi type I and there is a comoving perfect fluid having a linear equation of state $P=(\gamma-1)\rho$, $0<\gamma\leq 2$. We recall that in GR, there are two possible power-law solutions to the field equations when the geometry is Bianchi type I and the matter is a perfect fluid. If we assume \eqref{kasner} for the frame functions, then the two possible GR solutions are
\begin{enumerate}
\item $\gamma\not = 2$: $p=q=r=\frac{2}{3\gamma}$, $\rho=3p^2t^{-2}$ representing an isotropic $k=0$ Robertson-Walker solution.
\item $\gamma=2$: $p+q+r=1$, $p^2+q^2+r^2=1-2\rho_0$, $\rho=\rho_0t^{-2}$ where $\rho_0$ is a positive constant.  This solution represents a anisotropic Bianchi type I model containing a stiff fluid.
\end{enumerate}

As noted earlier, we must consider the two different scenarios for the solution to the spin-connection when considering Bianchi type I solutions in $F(T)$ teleparallel gravity.

\subsubsection{General Case: $M_1=M_2=M_3=0$}

In the general case, we have two different possible non-vacuum power-law solutions:
\begin{enumerate}
  \item $p=q=r$, $\rho=\rho_0 t^{-3\gamma p}$, $T=6p^2t^{-2}$ and $\rho_0=\frac{F_2}{2\kappa}(3\gamma p-1)\left({\sqrt{6}p}\right)^{3\gamma p}$.  Unlike the similar isotropic solution in GR, we see that the powers $p$, $q$ and $r$ are not fixed other than that they are equal.  This solution identically satisfies the first integrals obtained in equation \eqref{intFE} with $C_1=C_2=0$.  Further, the field equations restrict the form of the potential function
      \beq
      F(T)=F_1T^{\frac{1}{2}}+F_2T^{\frac{3}{2}\gamma p}.
      \eeq
      where $F_1$ and $F_2$ are arbitrary constants.

  \item $p+q+r=\frac{1}{\gamma-1}$, $\gamma \not = 1$, $\rho=\rho_0t^{-\frac{\gamma}{\gamma-1}}$, $T=(pq+pr+qr)t^{-2}$,
      and
      \beq
      \rho_0 = \frac{F_1}{2\kappa(\gamma-1)}(2pq+2pr+2qr)^\frac{\gamma}{2(\gamma-1)}.\label{rho1}
      \eeq
      where $F_1$ is a constant. Again the field equations fix the form of the potential function
      \beq
      F(T)=F_1T^{\frac{\gamma}{2(\gamma-1)}}.
      \eeq

      To compare with the similar GR case, we first note that we are no longer restricted to a stiff fluid ($\gamma\not =2$). We can also re-express equation \eqref{rho1} equivalently as
      \beq
      p^2+q^2+r^2 =\left(\frac{1}{\gamma -1}\right)^2 - 2\left(\frac{2}{F_1}(\gamma-1)\kappa\rho_0\right)^{\frac{2}{\gamma}(\gamma-1)},
      \eeq
      reminiscent of the Kasner relations \eqref{Kasner_relations}.  This solution represents an anisotropic Bianchi type I tele-parallel geometry containing a perfect fluid with linear equation of state
\end{enumerate}

\subsubsection{Special Solution  $M_1\not=0$, $M_2=M_3=0$}

Again we look only for solutions that do not reduce to constant torsion scalar or have a linear function of $F(T)$.  In the special case $M_1\not=0$, $M_2=M_3=0$ we obtain a solution
\begin{enumerate}
\item $r=1$ $p=q$, $L_1 = -\frac{(p-1)(2p(\gamma-1)+(\gamma-2))}{2p\gamma+(\gamma-2)}$ $\rho=\rho_0t^{-(2p+1)\gamma}$ where $\rho_0$ is a complicated expression involving $p$, $\gamma$ and $F_1$.  The potential function
\beq
F(T) = F_1 T^{\frac{(2p+1)\gamma}{2}}
\eeq
This solution represents an anisotropic Bianchi type I model containing a perfect fluid.  It is essentially the non-vacuum analog of the solution found earlier in the special case \eqref{spec_vac}.
\end{enumerate}


\section{Discussion} \label{sec:discussion}

In \cite{Coley:2022aty} space-times with a single affine symmetry, ${\bf X}$, were studied.  There, an algorithm was presented to find a general spin connection consistent with a single affine symmetry assumption.  The results in \cite{Coley:2022aty} easily transfer to problems in which the symmetry vectors commute, such as Bianchi I under study here. Employing the techniques developed in \cite{Coley:2022aty}, we began our investigation by determining the nature of the most general spin connection that can be employed in the analysis of Bianchi type I teleparallel geometries.  The most general solution respecting the Bianchi type I symmetries for the spin connection was computed.  We then determined what, if any, constraints are placed on the spin connection arising from the antisymmetric part of the $F(T)$ teleparallel gravity field equations.  That is, we determined consistent co-frame/spin connection pairs suitable to be used in the study of Bianchi type I teleparallel geometries.

We noted that in general, given that the torsion scalar contains six arbitrary functions, it may be possible to choose a spin connection such that the torsion scalar is identically constant. In this case the anti-symmetric part of the field equations are identically satisfied.  If $T$ is indeed a constant, then the symmetric part of the $F(T)$ teleparallel gravity field equations reduce to TEGR with a re-scaled gravitational coupling and a cosmological constant. This is not a general case.

Alternatively, if $T$ is not a constant, then we found that there exists essentially two classes of co-frame/spin connection pairs possible.  The first class is built from a co-frame/spin connection pair such that the corresponding proper co-frame (and consequently ``good tetrad'') is also invariant under the Bianchi I frame symmetries. In this first class, we obtain a completely general solution containing nine arbitrary functions of time before applying the symmetric part of the $F(T)$ teleparallel gravity field equations.  In this case the Lorentz transformation that builds the spin connection is a function of time only.

The second class of co-frame/spin connection pairs is constructed from a co-frame/spin connection pair such that the corresponding proper co-frame (and consequently ``good tetrad'') is no longer explicitly invariant under the Bianchi I frame symmetries.   Geometrically, this means that one of the basis elements is no longer aligned with the orbits of the symmetry. In the second class of co-frame/spin connection pairs, we have a special solution in that there are at most five arbitrary functions of time, and up to six arbitrary constants prior to solving the symmetric part of the $F(T)$ teleparallel field equations. In this special case, the Lorentz transformation that builds the spin connection is a function of one of the spatial coordinates.

The symmetric part of the $F(T)$ field equations are displayed for both classes. In the general case, or the first class, the spin connection does not enter into the symmetric part of the field equations.  Indeed, the field equations are exactly the same as if one was to start with a trivial spin-connection from the onset.  In the special case, or the second class, we clearly observe how one of the Lorentz parameters enters the field equations and the torsion scalar.  To our knowledge, this is the first time such a geometry has been presented within $F(T)$ teleparallel gravity theory.

In vacuum, we found that non-constant torsion scalar power-law solutions typically do not exist for general $F(T)$ teleparallel gravity.  We can construct power-law vacuum Bianchi type I solutions in two cases:
\begin{enumerate}
\item General case: The spin connection is constructed from the Lorentz parameter functions $A(t,x^j)=\alpha(t)$, $\Theta(t,x^j)=\vartheta(t)$, $B(t,x^j)=\beta(t)$, and $E(t,x^j)=\eta(t)$. The powers in the frame functions are all equal $p=q=r$ and the function $F(T)=F_0 T^{1/2}$. This solution is a $k=0$ TRW solution.
\item Special case: The spin connection is constructed from the Lorentz parameter functions $A(t,x^j)=\alpha(t)e^L_1x$, $\Theta(t,x^j)=\vartheta(t)+J_1x$, $B(t,x^j)=0$, and $E(t,x^j)=0$. The powers in the frame functions are $p=q$, $r=1$ and the function $F(T)=F_0 T^{p/2}$ ($p\not=2$).
\end{enumerate}
Interestingly, the assumption of power-law solutions in many cases implied that the resulting teleparallel gravity  theory had either $T=0$ or required that $F(T)$ be linear in $T$.  In such cases, $F(T)$ teleparallel gravity theory reduces essentially to GR with a re-scaled coupling constant and cosmological constant.  It is only in the two examples given above that a power-law vacuum solution results in a non-constant $T$ and a non-linear $F(T)$. In $F(T)$ teleparallel gravity, we find serious restrictions on the existence of power-law vacuum solutions. Of course there will exist power-law vacuum solutions in which the torsion scalar is constant (and when $F(T)$ is linear in $T$).

In the case of a perfect fluid with Bianchi type I geometry having a power law solution with a non-constant torsion scalar and non-linear potential function $F(T)$ we find a few possibilities:
 \begin{enumerate}
  \item General Case: ($\gamma \not = 1$)  The potential function has the form $F(T)=F_1T^{\frac{1}{2}}+F_2T^{\frac{3}{2}\gamma p}$.  The powers in the frame functions are all equal $p=q=r$ which means that the resulting solution is a $k=0$ TRW perfect fluid solution.

  \item General Case: The potential function has the form
      \beq
      F(T)=F_1T^{\frac{\gamma}{2(\gamma-1)}}.
      \eeq
      The powers in the frame functions must satisfy
      \begin{equation}
        p+q+r=\frac{1}{\gamma-1},\qquad p^2+q^2+r^2 =\left(\frac{1}{\gamma -1}\right)^2 - 2\left(\frac{2}{F_1}(\gamma-1)\kappa\rho_0\right)^{\frac{2}{\gamma}(\gamma-1)}
      \end{equation}
      This solution represents an anisotropic Bianchi type I tele-parallel geometry containing a perfect fluid with linear equation of state
\item Special Case:  The potential function has the form $ F(T) = F_1 T^{\frac{(2p+1)\gamma}{2}}$.  The powers in the frame functions must satisfy $r=1$ and $p=q$.  The parameter $L_1$ present in the spin connection must satisfy $$L_1 = -\frac{(p-1)(2p(\gamma-1)+(\gamma-2))}{2p\gamma+(\gamma-2)}.$$
    This solution represents an anisotropic Bianchi type I model containing a perfect fluid.
\end{enumerate}

This work illustrates some of the challenges in understanding affine frame symmetries and isometries in teleparallel gravity.  We have clearly shown that for a given co-frame, two different choices for the spin connection ansatz are allowable.  One choice for the spin connection is constructed from a Lorentz transformation that is function of time only.  The second choice involves constructing a spin connection from a Lorentz transformation that is a function of time and the spatial coordinates.  The metric and co-frame are in either choice the same. Both choices are valid, but choosing one over the other may lead to different behaviours in cosmological models. In the first choice, we are restricted to those Lorentz transformations that are invariant under the Bianchi I symmetry.  This does appear to be the general case.   However, we must be aware that there exists classes of spin connections which satisfy the symmetry requirements, but whose corresponding proper frame does not.

This analysis sets the foundation to determine explicitly co-frame/spin connection pairs suitable for broader classes of spatially homogeneous (Bianchi models) in $F(T)$ teleparallel gravity \cite{HMC}.


\section*{Acknowledgments}
AAC and RvdH are supported by the Natural Sciences and Engineering research Council of Canada. RvdH is supported by the Dr. W.F. James Chair of Studies in the Pure and Applied Sciences at St.F.X.

\section*{Author Declarations}

\subsection*{Conflict of Interest}
The authors have no conflicts to disclose.

\subsection*{Data Availability Statement}
Data sharing is not applicable to this article as no new data were created or analyzed in this study.


\section*{References}

\bibliographystyle{apsrev4-1}
\bibliography{../BIBFILES/Tele-Parallel-Reference-file}

\end{document}